\documentclass[a4paper,11pt]{article}
\pdfoutput=1 

\usepackage{jinstpub} 

\title{\boldmath Implementation of a Cost-Efficient Device for Wireless Photon Coincidence Detection}

\author[a,b]{E. Ipus,}
\author[a,b]{C. A. Melo-Luna,}
\author[a]{L. Giraldo,}
\author[b]{O. Vergara,}
\author[a,b,1]{J. H. Reina \note{Corresponding author.}}

\affiliation[a]{Centre for Bioinformatics and Photonics---CIBioFi, Edificio 320 No. 1069, Universidad del Valle, 760032 Cali, Colombia}
\affiliation[b]{
Departamento de F\'isica, Universidad del Valle, 760032 Cali, Colombia}

\emailAdd{john.reina@correounivalle.edu.co}

\abstract{State-of-the-art technology for pulse counter electronics  offers an important range of commercial devices, but such systems are usually expensive due to the complex logic used for this task. The use of  counting electronics in conjunction with photon counters can be used, for example, to perform experimental tests in Quantum Optics and Quantum Information Science.  Here,  we present the development and implementation of a low cost module for multiphoton coincidence statistics with detection windows of a few nanoseconds. The module consists of an array of logic gates, with a frequency operation of 250 MHz that corrects and amplifies the detectors signal.  The device characterisation was done by means of detection of Transistor-Transistor Logic (TTL) signals retrieved from a signal generator, and implemented in an optical setup.  The detected output signals (TTL pulses) were analysed and stored in a computer by means of a Field Programmable Gate Array (FPGA). Our module incorporates fundamental electronics that is currently used in the first experimental proof-of-principle tests in quantum information and molecular spectroscopy at CIBioFi.}


\arxivnumber{} 

\collaboration[c]{}

\proceeding{}

\begin{document}
\maketitle
\flushbottom
\section{Introduction}

\label{intro}
In less than three decades, a new technological revolution has been boosted by harnessing the fundamental principles of quantum mechanics, and  the superposition and entanglement of quantum states have been at the core of such a development. This, however,  implies a big challenge for testing and implementing remarkable protocols such as quantum computation, cryptography, and teleportation in quantum information science \cite{1,2,3,CAML2017}. 
In the case of optical technologies, the detection of individual pulses and photon-coincidences is fundamental \cite{7,8}, and setups for Light Detection and Ranging (LIDAR) \cite{Chevrier2017}, Fluorescence Correlation Spectroscopy (FCS) \cite{17, Becker}, molecular life-time emissions and quantum entanglement and correlations quantification \cite{Becker, S10,S12}, between others, rely  on this feature in a crucial way. Thus, photon statistics \cite{4,5,6} through the coincidence counting (detection of photons-simultaneously or within a small time window) provide a key ingredient in Science and Engineering.

Most of the time-dependent statistics of photons in different experimental schemes has been made possible thanks to a Time to Amplitude Converter (TAC) protocol, which  allows for the measurement of the time interval between incident pulses, from ``start'' to ``stop'', and generates an output pulse that is proportional to this time interval. This action  is commonly implemented in experimental setups used for  the characterisation of molecular systems
such as ``Time Correlated Single Photon Counting'' (TCSPC)  \cite{16, Becker02}.  This said, such a protocol for  coincidence counting  of multiple photons is nowadays expensive and easily saturable due to the long dead-time intervals around 8 MHz in reverse mode operation \cite{Lakowicz}, and even for 16 MHz \cite{Crotti2012}.

Time-to-Digital Converter (TDC) is another procedure that employs the ``start-stop'' principle. This can be entirely implemented on a  digital basis, thus avoiding the use of ADC (Analog-to-Digital) protocols that limit TDC applications to sub-micron technologies \cite{Lampton1994}. Current TDC technology could achieve a maximum frequency in the  5-10 GHz range and reach a maximum around 200 ps for the measurement accuracy \cite{Henzler}, and even a shorter sampling time \cite{malass2014, Zhang2017, Jiang2006}. Combining this technique with photon detectors such as the Avalanche Photodiode (APD), the sampling would be limited by the maximum rate allowed by the detector.  This method can be employed in diverse experiments, ranging from, e.g., materials surface reconstruction in the measure of the Time-of-Flight (ToF) of photons from a transmitter to a target and back to the detector \cite{Ivornicu2014}, fluorescence life-time imaging \cite{Field2014}, to TCSPC \cite{malass2014, Markovic2013}. Thus, a purely digital  proposal which offers an appropriate sampling rate at  a reduced  implementation cost \cite{8, 10, 9} can be very useful for the purpose here developed.
\begin{figure}
  \centering
 \includegraphics[width=12cm, height= 7cm]{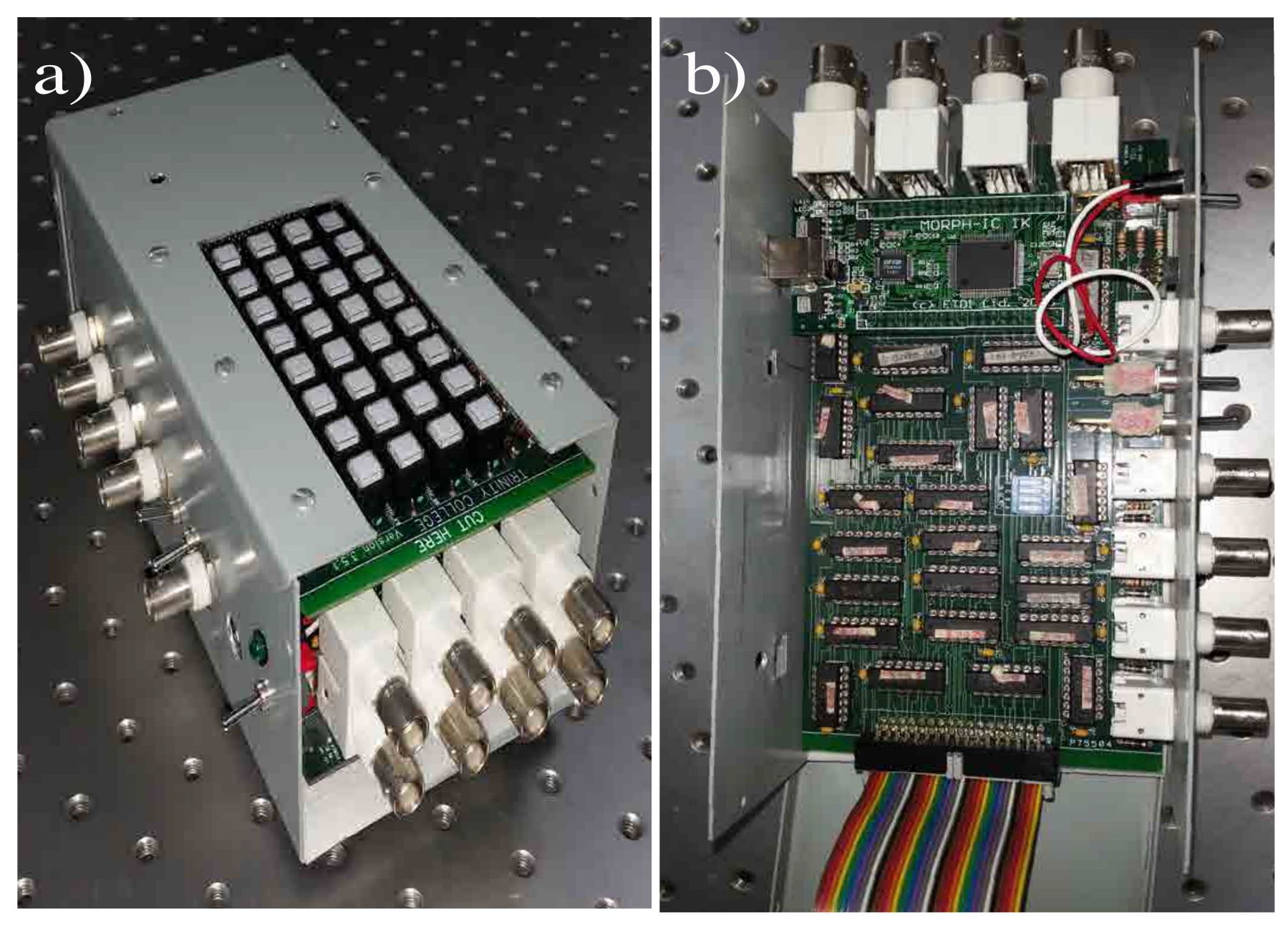}
\caption{\label{fig1}
Cards that exhibit the electronics of the first implemented  coincidence counting module in our laboratory based on  Branning's work \cite{8}: a) the electronics for the detection of the signals, and the final presentation of the coincidence counting module Branning's version. The $8\times4=32$ buttons on top show the 8 channels by 4 inputs that can be handled by the device. This ensures up to four-coincidences analysis in each available channel, b)  the structural location of the electronics logic that allow us the manipulation of the entry signals.
In this work, this module was expanded up to eight inputs and eight-fold coincidences. 
}
\end{figure}
\section{Experimental Development}
\subsection{Counting module assembly}
In 2009,  Branning {\it et al.} \cite{8} posed to change the TAC protocol by implementing a set of logic gates.  This assembly used a TTL pulse sent by a commercial photo counter to modify the pulse width, then defined the coincidence, and finally used a Peripheral Interface Controller (PIC) to count and store the data in a computer. After this, the PIC was replaced by a Field Programmable Gate Array (FPGA) that developed the same functionality as its predecessor but became better  suited and adaptable \cite{10}. This development is appealing due to the cost-efficiency of these devices, and besides, this also allows the scalability to 4$N$ inputs but using $N-1$ different Coincidences Counting Modules (CCM) \cite{9}.

We initially used a third generation proof board of Branning's CCM \cite{bran} and  assembled a device as shown in figure \ref{fig1}. 
This CCM was built on electronic boards of four layers each one. In one of them,  we located the different electronic devices, NANDs, ORs, and Multiplexors. In the second card,  the manual selection of input signals and the hosting of switches was configured, as can be seen in figure \ref{fig1}(a), and \ref{fig1}(b). Those devices are protected by a metallic box to preserve their electronic components, and to allow for an easier manipulation of the counting module.
The main limitation of this module, however,  is the reduced number of inputs (4) for the Single Photon Counting Modules (SPCM), given that, for example,   state-of-the-art quantum optics  experiments can entangle up to ten photons, and use around 20 APDs for their detection \cite{6}. 
In this sense, recent  efforts have expanded other counting strategies with different techniques that are able to reach up to 32 \cite{17}, and even 48 inputs \cite{15}.  

In this work, we report  the implementation of  a photon-counter coincidence module with a short response time (a few ns), and use as a counting device an FPGA DE0-nano model \cite{De0}.
In this module, we expand Branning  {\it et al.} initial proposal, by  increasing  the number of inputs up to 8 as well as the coincidences (8-fold).  In our implementation, we used integrated circuits of fast series (SN74FXXX) of different logical arrays, and a wireless module to communicate the data to a  software analyser.
The stages followed in the coincidence counting process are as follows: i) pulse shaping, ii) selection of the  input signal, iii) counting, and  iv) storing the acquired data. We next describe these steps.
\begin{figure}
  \centering
 \includegraphics[width=15.5cm, height= 5.5cm]{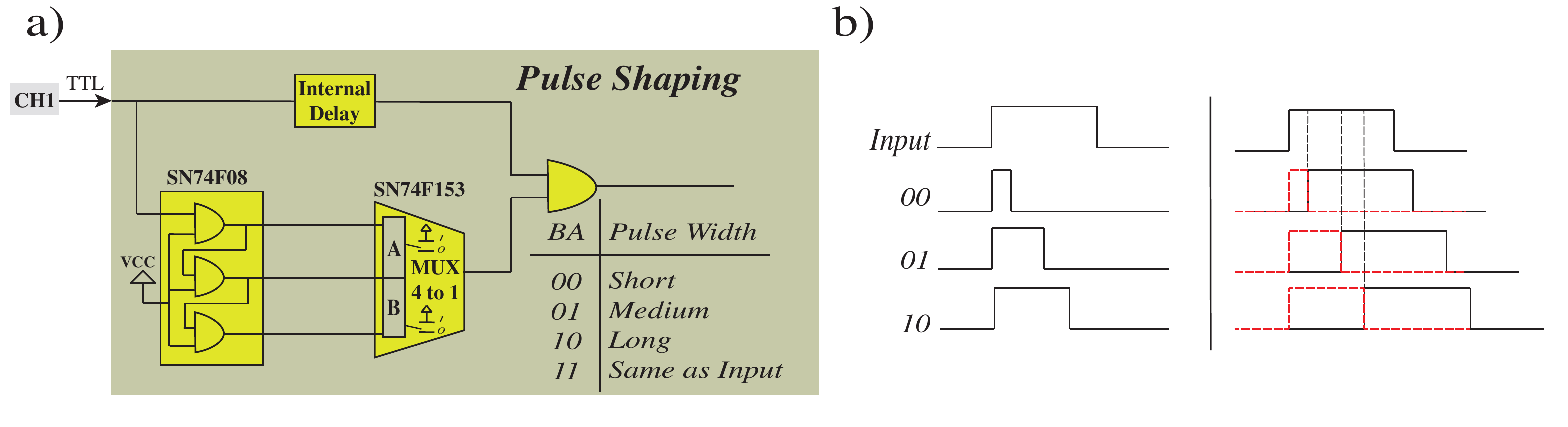}
\caption{\label{fig2}Schematical description of the pulse shaping process. a)  Stages followed by the input signal from CH1.  b) Sketch of the resulting signal in the lower path as a function of the $A,\,\,B$ status selectors.}
\end{figure}
\subsection{Shaping the input pulses}
A  description of the pulse shaping process is schematically given in figure \ref{fig2}(a). First, the signal splits into two equal and temporal synchronised signals, next both of them are delayed but subjected to the following two-path rule. In the upper path, the delay is controlled through a double inverter logic circuit which is equivalent in time to the application of two digital gates. In the lower path, the delay is defined by the selectors $A$ and $B$, where ``$0$'' and ``$1$'' denote off and on states, respectively. This delay  finally defines the width of the output pulse, as is shown in figure \ref{fig2}(b). 
\begin{figure}
  \centering
 \includegraphics[width=9cm, height= 6cm]{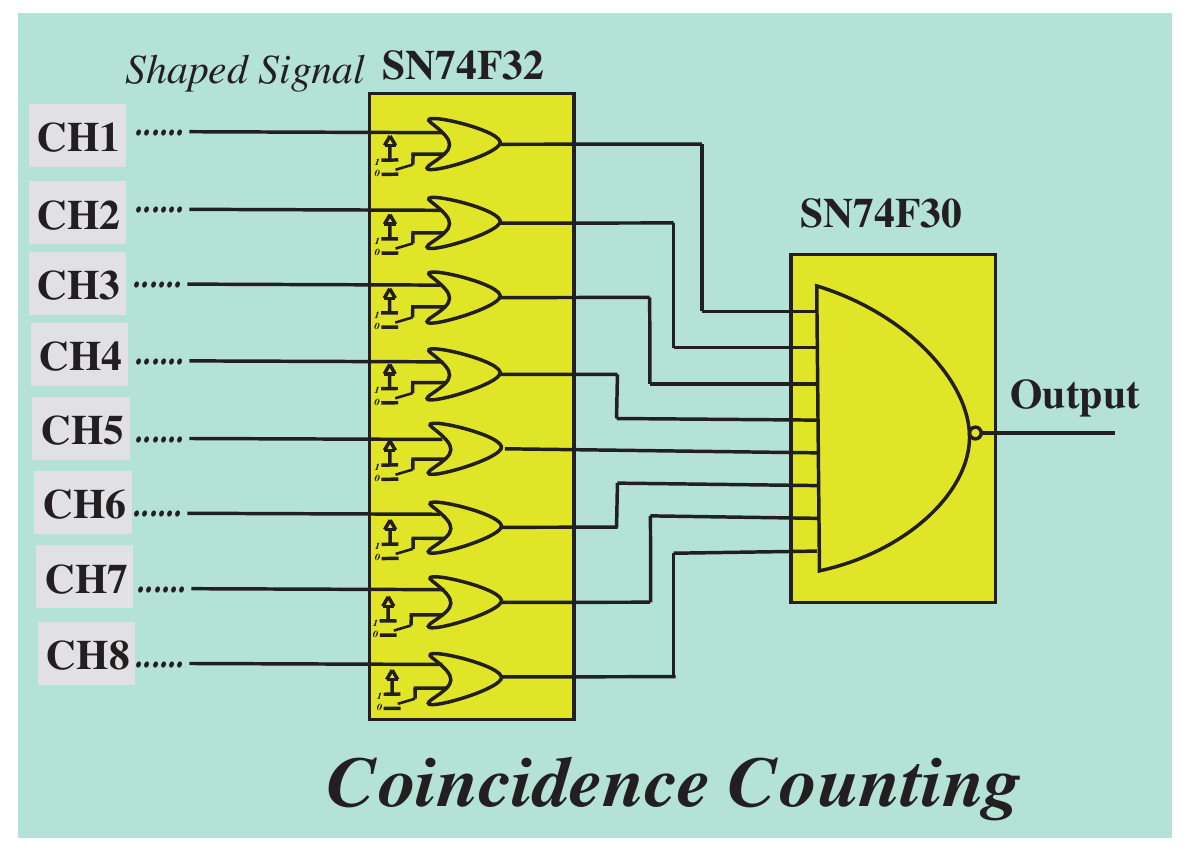}
\caption{\label{fig3}Scheme of signals selector: OR-gates (SN74F32) are applied on each shaped pulse, and over the logical state of switches manually defined. Only the selected signals get involved in the coincidence process performed by an AND-gate with multi-inputs (SN74F30).}
\end{figure}

\subsection{Selecting the input signal}
All the input pulses are shaped as explained before but only the modules' user defines the number of inputs that are to be considered in the coincidence process.  The selection of  the signals is implemented via a switching system that allows  the following: to select a channel (or channels) for detection and counting the pulses, and then, counting the coincidences between signals.  Each shaped input is compared with the logic state of a switch (``0'' or ``1'') through an OR-gate (SN74F32), and the logic result is transmitted to the coincidence channel performed by an AND-gate (SN74F30), as is sketched in the figure \ref{fig3}, to be finally  counted.  The coincidence protocol is just the logical comparison of the selected pulses in a defined time interval (a few ns) through an AND-gate. To sum up,  the definition of a coincidence is limited by the electronics logical response time at the gate, and by the number of  followed buffers. 
\begin{figure}
  \centering
  \includegraphics[width=13cm]{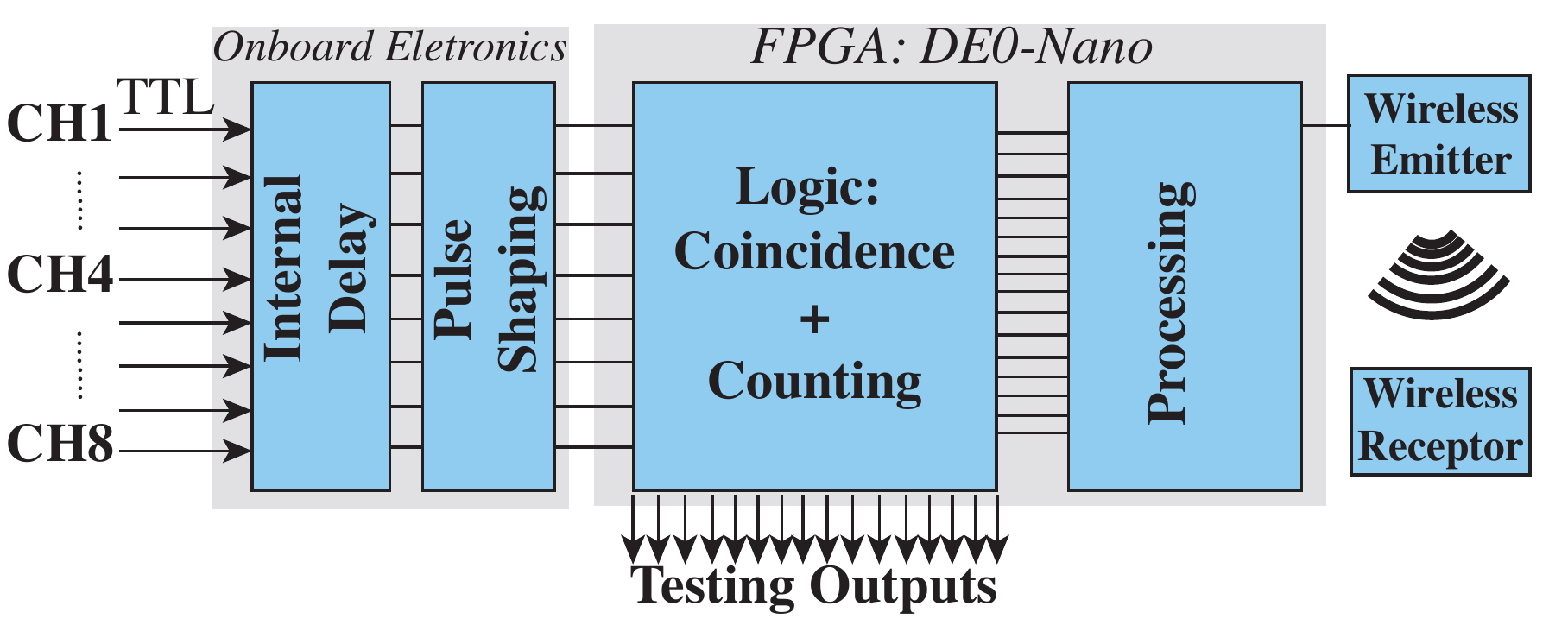}
\caption{\label{fig5}Scheme of the input processing and data transmission protocols. Each input signal is delayed, shaped, and counted. If the selection process defines more than one input, then the coincidence will be counted and saved. These data were collected during a defined period,  the so-called  {\it integration time},  $2\, \mu\text{s} \leq \tau_{count}\leq 1$ s, and were finally sent through an intercommunication wireless port to a  receptor module by packages that were stored in a computer.}
\end{figure}

\subsection{Counting, saving and data acquisition}
For this purpose, the final pulses are counted and stored for later acquisition in readable files in a computer. This is developed by means of an  FPGA of programmable logic and connections under  \textit{VHSIC\footnote{Acronym for Very High Speed Integrated Circuits} Hardware Description Language (VHDL)}.  A MORPH-1C-II system \cite{IC}  has been recently introduced, in four channels protocols \cite{bran}, due to its Multi Protocol Synchronous Serial Engine that allows programming and re-programming by means of a USB port in a fraction of a second, i.e.,  of at least $0.2$ s \cite{11}. 
This implementation, however, has a practical drawback since it requires programming every time that a lack of power supply arises and, operatively speaking, a plug-\&-play usage with this kind of FPGA is not possible. Another hurdle arises from this model's processing capability due to the fact that an increment in the number of possible input signals implies an increment of the number of selecting switches, as shown in figure \ref{fig1}(a) and figure \ref{fig3}. As an alternative, this switching system can be included in a software interface using the logical elements of the FPGA  instead of hardware buttons. In the latter, an FPGA with a larger number of logical elements is to be required.

 In our implementation, we resort to an FPGA with a ROM memory for the counting process as the DE0 nano \cite{De0} from the Cyclone\texttrademark IV family \cite{11}; this device has two headers of 40 pins each one, where 72 pins are input/output, 4 power pins, and 4 GND pins; connection to a USB port for its output signals and 23 pins for connection to a JTAG interface (standard protocol to develop debugging tools \cite{12}). Another key feature of this model is the maximum acceptable frequency of $153\,\text{MHz}$ for the input signals, and a larger  number of logical elements: $22.320$ compared to $4.608$ of the MORPH-1C-II. These features allow the detection of different physical systems that range from the single photon counting and coincidences through to the radioisotopes in positron emission tomography \cite{18}.

For programming this FPGA, we used the  compilation environment Quartus\texttrademark. The particular code used here can create eight registers of independent counts of 16 bits available each one to define the  channel of the register. This number of channels stems from the module inside that compares the selected input signals between them to count coincidences in the FPGA.  To obtain the  saved data, we used XBee modules that allow the wireless communication between the FPGA and a designated  computer \cite{13}. This process relies on the time needed to send the data by the transmission module ($\tau_{send}$). In order to overcome lacks in the processing, it is useful to define integration times to count in the FPGA in the same order ($\tau_{count} \sim \tau_{send}$). The required variables  in the process of wirelessly data sending are the \textit{sending  rate} (number of data send per second) and the \textit{bit length} (``size'' of the data), which depend on the number of channels to count. These variables determine the file transmission and reception, and are set in the code that is implemented within the FPGA and the XBee (X$-$CTU) module.  Finally, the file arrives to a reception module that is connected to a  computer to visualise and analyse the counts, as schematically shown in figure \ref{fig5}.
\begin{figure}
  \centering
  \includegraphics[width=9cm]{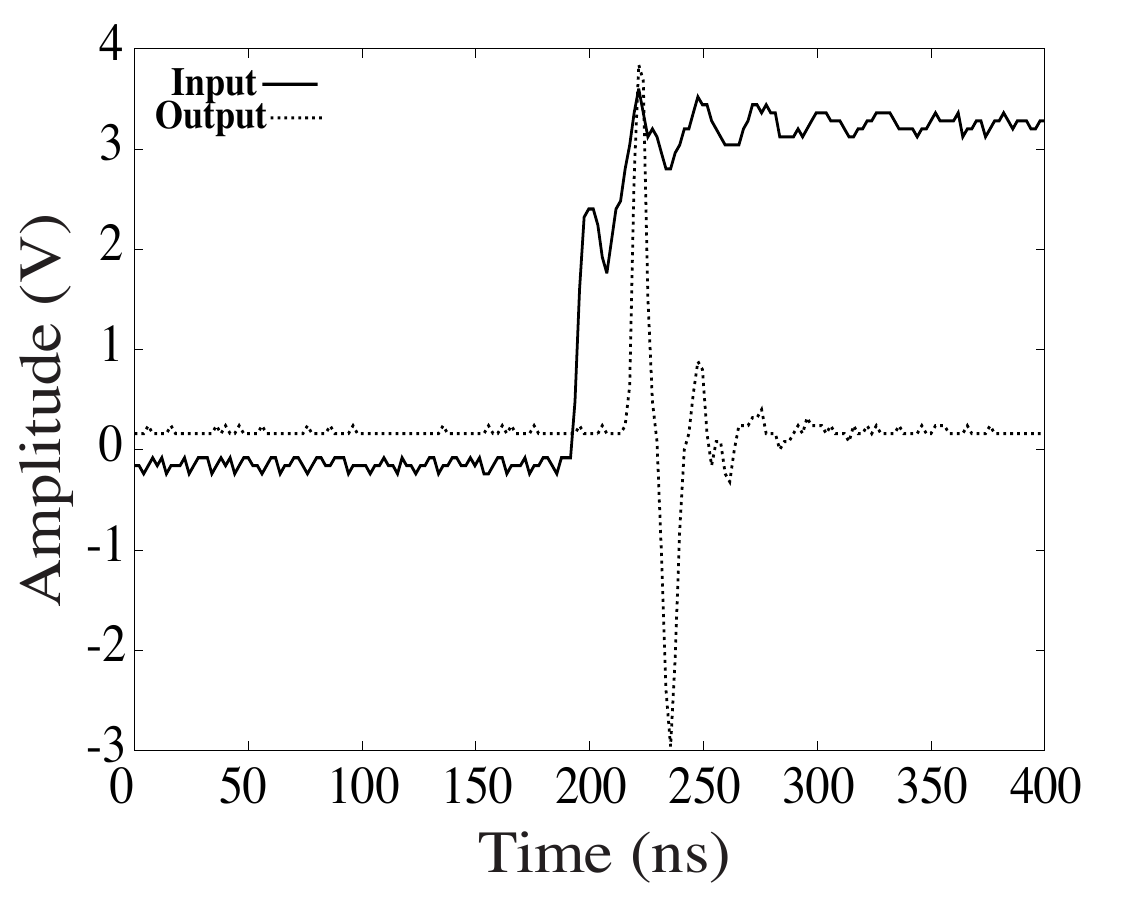}
\caption{\label{fig6}Comparison between an input signal of $10\,\text{KHz}$ and the output response of the configuration ``$00$'' during  the pulse shaping stage.}
\end{figure}
\subsection{Device estimated cost}
The total cost involved in the construction of the coincidence counting device can be estimated as follows: i) FPGA DE0-nano $\sim 190$ USD, ii) PCBs (assembling cards) $\sim 50$ USD, iii) the complete set of fast series circuits (SN74FXXX)  $\sim 30$ USD, iv) additional electronics and housing (switches, cables, LEDs, BNC connectors, metallic box) $\sim 200$ USD. These give an approximate total of about $470$ USD. This is to be contrasted with typical commercial devices available in the market (with similar features) that can be well above $2500$ USD (e.g., Canberra Model 2040 and National Instruments modules). 
\begin{figure}
  \centering
  \includegraphics[width=\textwidth]{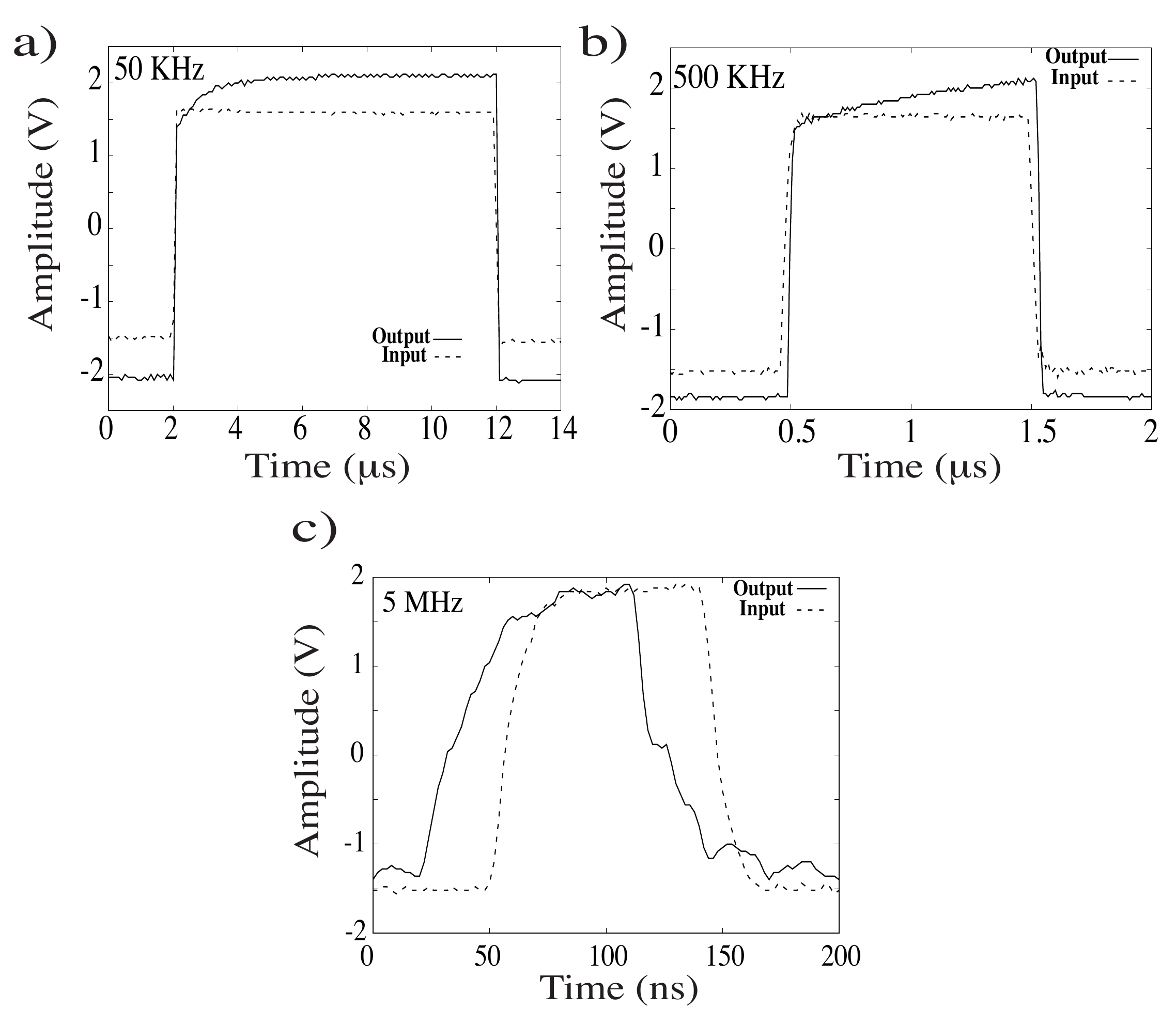}
\caption{\label{fig7} Input signals of different frequencies in the  ``11'' configuration: a) the $50\, \text{KHz}$ signal presents a rounded corner due to the amplification response of the pulse shaping stage: the time-scale of the pulse width is enough to see  the maximum amplification of about $0.5\,\,V$ between the output and the input signals, b) in the $500\,\text{KHz}$ pulse, the amplification process is also reached, but this exhibits a cut-off due to the increase of the repetition frequency, and c) an input pulse of $5\,\text{MHz}$ is used in the module: the amplification is not evident and an apparent time shift exhibited  by the output signal at FWHM around $\sim 16$\% of its width takes place. This is attributed  to the usage of a  trigger signal  that did not allow for a proper compensation of the output signal.}
\end{figure}
\begin{figure}
  \centerline{\includegraphics[width=10cm]{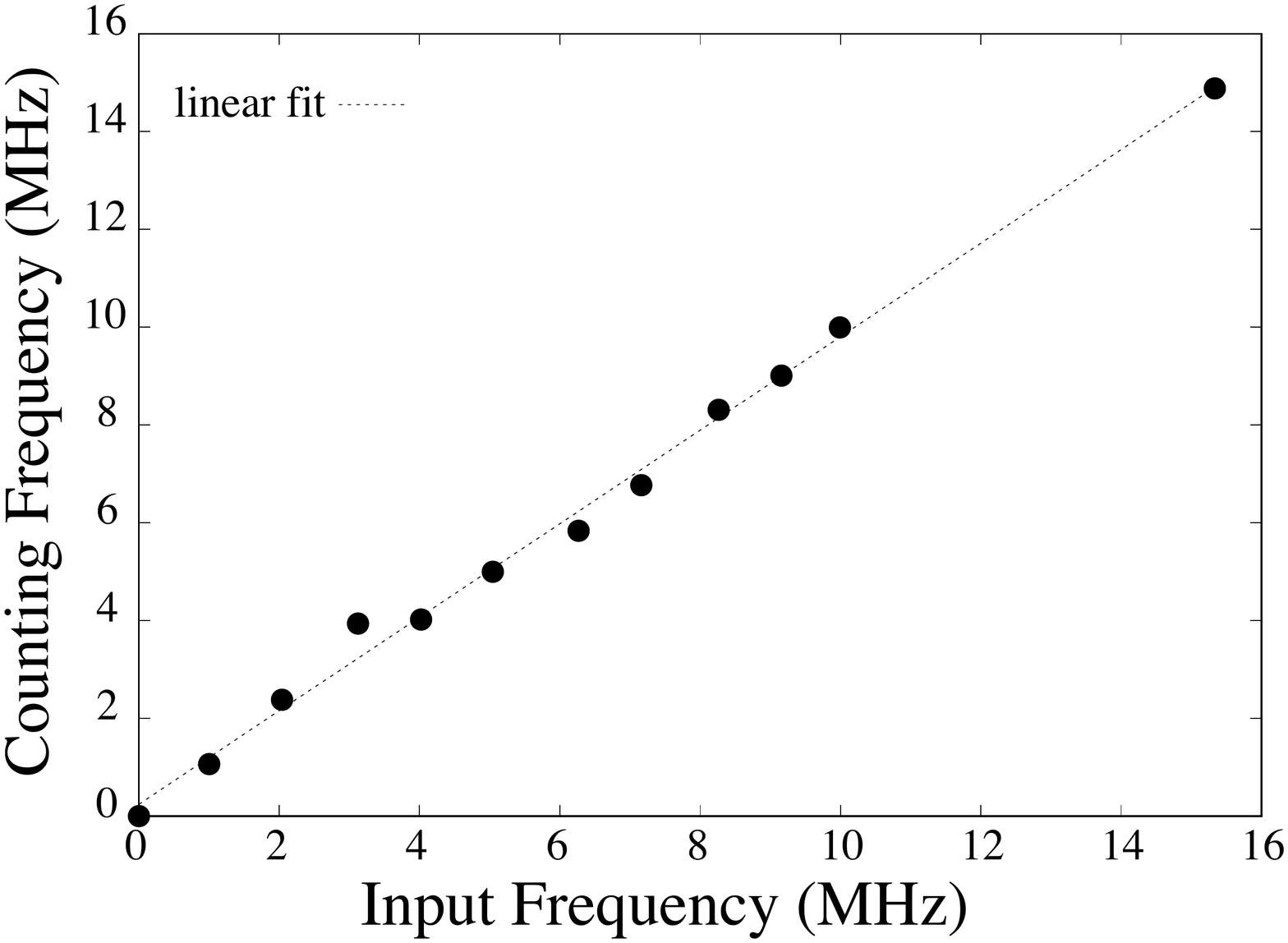}}
\caption{\label{fig8}Single channel counting rate for the  ``$11$'' configuration. The functional dependence  is linear, with calculated slope  $m=0.9561\pm0.0213$. An almost identical result was obtained   for the ``$00$'' configuration (not shown).}
\end{figure}
\section{Device Response and Discussion}

The coincidence counting module was carefully checked in each one of the previously mentioned stages. The pulses used for testing this module were generated by a square pulse generator set at three different frequencies (50 KHz, 500 KHz, and 5 MHz). As mentioned before, the input pulses can be modified in their temporal widths by using different configurations of the selectors in the MUX of the pulse shaping stage. In this sense, the short width of the pulse is reached in the configuration ``$00$'', and the minimal modification over the input signal should be obtained  in the ``$11$'' configuration. 

In  figure \ref{fig6} we show the effect of the pulse shaping in the configuration ``$00$''  over a square input signal in the counting module. Here, we observe that the output exhibits a damped oscillator-like behaviour,  possibly due to the response time at the logic circuits.  This input signal has a frequency of $10\, \text{KHz}$, which means an average width of $\sim 5\,\text{ms}$, and the shaping process produces an output of $\sim 15\,\text{ns}$; this represents a reduction of at least five orders of magnitude of the former width. Although the plot shows a delay of a few nanoseconds between the input and the output signals, this difference is comparably small: $\sim 10^{-4}$\, \% of  the input pulse width. 
 We also verified the output signal in the configuration ``$11$'', and were able to find an increment in the amplitude when compared to  the input pulse, as can be seen in  figure \ref{fig7}.  

This could be achieved due to the amplification performed by the shaping electronics at this stage and the impedance of the cables used for this experiment. However, this amplification process takes a time comparable to the time scale of these inputs to reach  a maximum value. This behaviour is responsible for the `round corner' in the amplification output in figure \ref{fig7}(a). When the frequency is changed, say by one order of magnitude, the shaping, initially rounded, is modified to a triangular-like shape, as shown in figure \ref{fig7}(b).  
This means that the amplification procedure is slower than the period of the input signal, and the maximum amplitude is not reached before the temporal width vanishes. In addition, frequencies in the range of MHz exhibit an alteration  of the output pulse and a ``non-evident''  voltage amplification process. 
For the case of a  5 MHz input signal, an apparent earlier appearance of the output (with respect to the input) signal is observed in figure \ref{fig7}(c). We estimate the uncertainty due to this time shift presented by the output signal at FWHM around $\sim 16$\% of its width. This is due to the usage of an oscilloscope's trigger signal  that did not allow us to  produce a proper compensation of the output signal.

This visualisation issue, however, does not compromise the functioning of the counting device, since the amplifying response during the shaping stage does not affect the efficiency of the counting process due to the fact that the counting frequency is proportional to the input frequency, as shown in figure \ref{fig8}.  We found that, in the configuration ``$11$'', the frequency of the input signal is proportional to the counting frequency by a constant value $0.9561\pm0.0213$. The ``$00$'' configuration exhibits a likewise proportionality with a very similar constant.

\section{Concluding Remarks}

We have built a cost-efficient (compared to commercial prices)  counting module device of eight  inputs, as well as eight-fold coincidence channels that exhibit key features that can be used in different areas of science, engineering and the medical sciences, and in particular, in quantum and non-linear optics.  These features allow us to perform a more in depth analysis about incident photons in quantum optics and quantum information experiments.  The reported counting module has a response time of a few nanoseconds and works for incident signals with frequencies up to 150 MHz. This  module is currently being implemented in our laboratory for the detection of Werner-like states which have  recently been proposed as a novel resource for  quantum game strategy in a protocol that  requires neither quantum  entanglement nor nonlocality  as a resource \cite{CAML2017}.  
We work on the FPGA programming to improve the module for data transfer in quantum tomography of one and two polarisation photonic qubits experiments.  

The device here implemented for photon coincidence detection shall also be used in quantum interferometry, and photo-luminiscence detection in molecular spectroscopy experiments.

\acknowledgments
E.I. thanks the ``Virginia Gutierrez de Pineda Young Researcher and Innovator'' COLCIENCIAS programme. We would like to thank Universidad del Valle (grant  7930), COLCIENCIAS (grant 71003), and Fondo CTeI-Sistema General de Regal\'ias (contract BPIN  2013000100007) for  financial support.

\bibliographystyle{JHEP}
\bibliography{ccm}








\end{document}